\documentclass[preprint2]{aastex}

\newcommand {\hii}{H\,{\sc ii}} 
\newcommand {\kms}{\relax \ifmmode {\,\rm km\,s}^{-1}\else \,km\,s$^{-1}$\fi}
\newcommand {\ha}{\rm H$\alpha$}

\newcommand {\oiii}{[O\,{\sc iii}]}
\newcommand {\nii}{[N\,{\sc ii}]}
\newcommand {\sii}{[S\,{\sc ii}]}
\newcommand {\hd}{HD\,5980}

\shorttitle{NGC\,346}
\shortauthors{Naz\'e et al.}

\received{2002 May 31}
\begin{document}

\title{An X-ray investigation of the NGC\,346 field in the SMC (1) : the LBV \hd\ and the NGC\,346 cluster}

\author{Y. Naz\'e\altaffilmark{1,8}, J.M. Hartwell\altaffilmark{2},
I.R. Stevens\altaffilmark{2}, M. F. Corcoran\altaffilmark{3},
Y.-H. Chu\altaffilmark{4}, G. Koenigsberger\altaffilmark{5},
A.F.J. Moffat\altaffilmark{6}, V.S. Niemela\altaffilmark{7}} 

\altaffiltext{1}{Institut d'Astrophysique et de G\'eophysique,
Universit\'e de Li\`ege, All\'ee du 6 Ao\^ut 17, Bat. B5c, B 4000 -
Li\`ege (Belgium); naze@astro.ulg.ac.be} 
\altaffiltext{2}{School of Physics \& Astronomy, University of
Birmingham, Edgbaston, Birmingham B15 2TT (UK); jmh@star.sr.bham.ac.uk,
irs@star.sr.bham.ac.uk} 
\altaffiltext{3}{Universities Space Research Association, High Energy
Astrophysics Science Archive Research Center, Goddard Space Flight
Center, Greenbelt, MD 20771; corcoran@barnegat.gsfc.nasa.gov} 
\altaffiltext{4}{Astronomy Department, University of Illinois, 1002
W. Green Street, Urbana, IL 61801; chu@astro.uiuc.edu} 
\altaffiltext{5}{Instituto de Astronom\'{\i}a, Universidad Nacional
Aut\'onoma de Mexico, Apdo. Postal 70-264, 04510, Mexico D.F. (Mexico);
gloria@astroscu.unam.mx} 
\altaffiltext{6}{D\'epartement de physique, Universit\'e de Montreal,
C.P. 6128, Succ. Centre-Ville, Montreal, QC, H3C 3J7 (Canada);
 moffat@astro.umontreal.ca} 
\altaffiltext{7}{Facultad de Ciencias Astron\'omicas y Geof\'{\i}sicas,
Universidad Nacional de la Plata, Paseo del Bosque S/N, B19000FWA La
Plata (Argentina); virpi@fcaglp.fcaglp.unlp.edu.ar} 
\altaffiltext{8}{Research Fellow F.N.R.S.}

\begin{abstract}
We present results from a {\it Chandra} observation of the NGC\,346 
cluster. This cluster contains numerous massive stars and is responsible for 
the ionization of N66, the most luminous \hii\ region and the largest 
star formation region in the SMC. In this first paper, we will
focus on the characteristics of the main objects of the field.
The NGC\,346 cluster itself shows only relatively faint X-ray emission 
(with $L_X^{unabs}\sim 1.5\times 10^{34}$~erg~s$^{-1}$), tightly 
correlated with the core of the cluster. In the field also lies \hd, 
a LBV star in a binary (or possibly a triple system) that is detected for the first 
time at X-ray energies. The star is X-ray bright, with an unabsorbed 
luminosity of $L_X^{unabs}\sim 1.7\times 10^{34}$~erg s$^{-1}$, but 
needs to be monitored further to investigate its X-ray variability over a 
complete 19\,d orbital cycle. The high X-ray luminosity may be associated
either with colliding winds in the binary system or with
the 1994 eruption. \hd\ is surrounded by a region of diffuse X-ray
emission, which is a supernova remnant. While it may be only a chance 
alignment with \hd, such a spatial coincidence may indicate that the
remnant is indeed related to this peculiar massive star. 
\end{abstract}


\section{Introduction}

The Small Magellanic Cloud (SMC) is an irregular galaxy at a distance
of 59~kpc \citep{ma86} that forms a pair with the Large Magellanic Cloud
(LMC). Both are satellites of our own Galaxy. The interstellar
extinction towards the Magellanic Clouds is low, allowing studies of the
X-ray sources to be undertaken with clarity. \\

Previous X-ray observations of the SMC have been made with the {\it
Einstein Observatory}, {\it ASCA} and {\it ROSAT}. These observations
included surveys of the point source population
\citep{ka99,sc99,hfp,sa00,yo00} and in particular studies of the
properties of the X-ray binary population \citep{ha00}. The recent launch of
{\it Chandra}, however, provides an opportunity to study the SMC with a
far greater sensitivity and spatial resolution than ever before. \\

We have obtained a Chandra observation of the giant \hii\
region N66 \citep{he56}, the largest star formation region
in the SMC, in order to study the young cluster NGC\,346 and
its interaction with the surrounding interstellar medium. 
This cluster contains numerous massive stars \citep{ma89}, 
of spectral types as early as O2 \citep{wal02}. The large number of 
massive stars is not the only feature of interest in this field. On the 
outskirts of NGC\,346 lies the remarkable star \hd, which underwent 
a Luminous Blue Variable (LBV)-type eruption in 1994. This massive 
binary (or triple?) system 
has been monitored and analysed for its varying spectral and photometric 
properties in visible and UV wavelengths since the early 80's 
(Koenigsberger et al. 2000; Sterken \& Breysacher 1997, and 
references therein). Previous X-ray observations of NGC\,346 
\citep{ikt,hfp} detected a bright extended
source around \hd\ which has been attributed to a supernova
remnant (SNR); however, the crude instrumental resolution
prohibited unambiguous detections of a point source associated
with \hd.  In addition to \hd, the young massive stars of NGC\,346 
and their interacting winds are likely to
produce X-ray emission, but have not been detected previously.
The sensitive, high-resolution {\it Chandra} observation of the
NGC\,346 field thus provides an excellent opportunity to study a
variety of phenomena involving emission at X-ray energies. \\

In this paper, we will describe in \S~2 the observations used in this 
study, then discuss the available data on NGC\,346, \hd, and the extended 
emission in \S~3, 4, and 5, respectively. Finally, conclusions are 
given in \S~6. The second part of this work (Naz\'e et al., paper II) 
will describe the properties of the other sources detected in the field.\\

\section{The Observations and Data Analysis}

\subsection{X-ray Observations}

NGC\,346 was observed with {\it Chandra} for the 
XMEGA\footnote{http://lheawww.gsfc.nasa.gov/users/corcoran/xmega/xmega.html} 
consortium on 2001 May 15--16 for 100 ks (98.671 ks effective, ObsID = 1881,
JD$\sim$2\,452\,045.2d).  The data were recorded by the ACIS-I0, I1, I2,
I3, S2 and S3 CCD chips maintained at a temperature of
$-120^\circ$C. The data were taken with a frame time of 3.241s in the
faint mode. The exposure was centered on the cluster, with \hd\ lying
1.9\arcmin\ to the north-east of the aimpoint on ACIS-I3. Our faintest
sources have fluxes of about $2\times 10^{32}$~erg~s$^{-1}$ (see
paper II), assuming a distance of 59~kpc for the SMC.\\

We excluded bad pixels from the analysis, using the customary bad-pixel
file provided by the {\it Chandra} X-ray Center (CXC) for this
particular observation. We have searched the data for background flares,
which are known to affect {\it Chandra} data, by examining the
lightcurve of the total count rate, but no flares were found. Event Pulse 
Invariant (PI)
values and photon energies were determined using the FITS Embedded File
(FEF) acisD2000-01-29fef\_piN0001.fits.  \\

For long exposures, removing the afterglow events can adversely affect
the science analysis (underestimation of the fluxes, alteration of the
spectra and so on)\footnote{see caveat on
http://asc.harvard.edu/ciao/caveats/acis\_cray.html}. We thus computed a
new level 2 events file by filtering the level 1 file and keeping the
events with {\it ASCA} grades of 0, 2, 3, 4 and 6, but without applying
a status=0 filter. Throughout this paper, we will use this new file for
all scientific analysis.\\

Further analysis was performed using the CIAO v2.1.2 software provided
by the CXC and also with the FTOOLS tasks. The spectra were analysed and
fitted within XSPEC v11.0.1 \citep{ar96}. In Fig. \ref{chandracol}, we 
show a 3 color image of the Chandra data of the NGC\,346 cluster, \hd\
and its surroundings. We will discuss the features of this image further 
in later sections. \\

We have investigated problems due to Charge Transfer Inefficiency 
(CTI). To remove these CTI effects, we used the algorithm of 
\citet{to00}. Checking several sources on different 
chips and/or at different positions on each CCD, we concluded that 
the impact of the CTI was small and that it did not change significantly 
the spectral parameters found by using non-corrected data. 
When a difference was apparently found, we discovered that the best fits
found on corrected and non-corrected data gave similar $\chi^2$. 
As the work of \citet{to00}is not yet an official CXC product\footnote{http://www.astro.psu.edu/users/townsley/cti/}, 
we chose to present here only the results of the non-corrected data. \\

\subsection{{\it HST} WFPC2 Images}

Two {\it HST} WFPC2 observations of the N66 / NGC\,346 region in \ha\
are available from the STScI archives. They were taken on 1998 Sept.~27
and 2000 August 7 for the programs 6540 and 8196, respectively. \hd\ is
centered on the PC in the observations from program 6540 (see Fig. 
\ref{hst6540x}), while program
8196 shows an area situated further south. The observations were made
through the $F656N$ filter for 2070s (Prog. \# 6540) and 2400s (Prog. \#
8196). This filter, centered at 6563.7 \AA\ with a FWHM of 21.4\AA,
includes the \ha\ line and the neighboring [N~{\sc ii}] lines.  \\

The calibrated WFPC2 images were produced by the standard {\it HST}
pipeline. We processed them further with IRAF and STSDAS routines.  The
images of each program were combined to remove cosmic rays and to
produce a total-exposure map. No obvious correlation between X-rays 
features and the \ha\ emission is seen in the {\it HST} data. 
Fig.~\ref{hst6540x} will be discussed more extensively in \S~5.\\

\section{The NGC\,346 Star Cluster}

The NGC\,346 cluster has a large number of massive stars,
containing almost 50\% of all early-type O stars in the SMC
\citep{ma89}.  It is responsible for the ionization of N66, 
the most luminous \hii\ region in the SMC \citep{ye91}.
However, the NGC 346 cluster has not been
detected in X-rays until this {\it Chandra} observation.  Even
though NGC\,346 lies near the joint of the four ACIS-I CCD
chips and part of the cluster is lost in the gaps between
CCDs, the X-ray emission from the core of NGC\,346 is
unambiguously detected in the corner of the ACIS-I3 chip.
The X-ray emission appears extended with multiple peaks
corresponding to the bright blue stars MPG 396, 417,
435, 470, and 476 (Fig. \ref{dssmpg}).  As only $\sim$300 counts are detected
and the X-ray peaks are not well-resolved, we will analyze
only the overall emission from the cluster.\\

The total count rate of this emission in the $0.3-10.0$~keV band is
$3.04\pm 0.36\times 10^{-3}$~cnts~s$^{-1}$. 
The spectrum of the NGC\,346 cluster was extracted from a square 
aperture of $\sim$45\arcsec$\times$45\arcsec\ with a carefully
chosen background region, as close as possible to the cluster. As we
regard the cluster X-ray emission to be extended, we have used the
{\it calcrmf} and {\it calcarf} tools to generate the appropriate
response files. The best fit ($\chi^2_{\nu}$=0.90, $N_{dof}$=50, $Z=0.1
Z_{\odot}$) to this spectrum was an absorbed {\em mekal} model \citep{ka92}
with the following properties (see Fig.~\ref{specclus}):
an absorption column density $N(H)$ of $0.42_{0.17}^{0.83}\times 
10^{22}$~cm$^{-2}$ and a temperature $kT$ of $1.01_{0.61}^{2.23}$~keV. 
The luminosity in the $0.3-10.0$~keV band
was $5.5\times 10^{33}$~erg~s$^{-1}$, which corresponds to an unabsorbed
luminosity of $1.5\times 10^{34}$~erg~s$^{-1}$. The absorption column
is comparable to the value expected for NGC\,346 (see paper II), but
the temperature is one of the lowest we found, except for the extended
source around \hd.\\

In the region of X-ray emission in NGC\,346, 30 blue stars were 
detected by
\citet{ma89}: MPG 435, 342, 470, 368, 476, 396, 487, 417, 370, 471, 467,
495, 451, 330, 455, 454, 500, 445, 468, 375, 508, 395, 439, 371, 485,
561, 374, 366, 557, 400 (by increasing magnitude). Of these, 16 have
known spectral types that we can use to convert magnitudes to bolometric
luminosities. The bolometric correction factors were taken from
\citet{hum}. We can then compute the expected X-ray luminosities using
$log(L_X^{unabs})=1.13\log(L_{BOL})-11.89$ \citep{ber}. If the X-rays
were coming only from these 16 stars, we should expect a total
unabsorbed luminosity of $2\times 10^{33}$~erg~s$^{-1}$. This represents
only 13\% of the detected luminosity. Several hypotheses could explain
this discrepancy: a metallicity effect (Bergh\"ofer's relation was 
determined for Galactic stars but SMC stars have weaker winds);
X-ray emission from the other (unclassified) stars in this region, 
especially from the low-mass stars (see e.g. Getman et al. 2002); 
additional emission linked to 
interactions in binary systems (such as colliding winds);  or an 
extended region of hot gas as the winds from the individual stars 
combine to form a cluster wind \citep{oz97,canto}.\\\\

{\it Chandra} has also observed X-ray emission from other very young
(age $<5$~Myr) stellar clusters, comparable to NGC\,346. The Arches
cluster in the Galactic Center region has been observed to have
emission from several different regions, with a total X-ray luminosity
of $L_X^{unabs}\sim 5\times 10^{35}$~erg~s$^{-1}$ \citep{yz02}. NGC\,3603 has
also been observed with {\it Chandra} \citep{moff02} and the total
unabsorbed cluster luminosity is $L_X^{unabs}\sim 1\times 10^{35}$~erg~s$^{-1}$
(though in this case it is clear that a substantial fraction of the
emission is from point sources in the cluster). For both the Arches and
NGC\,3603 clusters the fitted temperature of the X-ray emission is
higher than the $kT\sim 1$~keV value for NGC\,346.\\

It is important to note that the calculations presented by \citet{canto} 
and \citet{ra01} for the Arches cluster assume that there are about 
60 stellar sources having mass-loss rates of 10$^{-4}$ M$_{\odot}$ 
yr$^{-1}$.  If we take the list of the bluest stars in NGC 346 given 
by \citet{ma89}, the first 60 sources have spectral types in the range 
O3-B0.  Assuming mass-loss rates for 8 SMC stars given by \citet{pu96} 
to be typical, then all stars 
in NGC\,346 (excluding \hd) have  mass-loss rates smaller than 10$^{-4}
$ M$_{\odot}$ yr$^{-1}$. Only MPG 435 (O4III(n)(f)) has a mass loss rate 
as high as 9$\times$10$^{-5}$ M$_{\odot}$ yr$^{-1}$. However, the majority 
of these 60 stars are late O-type Main Sequence stars or cooler, which 
must have much smaller mass-loss rates (10$^{-6}$ M$_{\odot}
$ yr$^{-1}$ or less), and they are more widely
distributed than in Arches.  Thus, it is not surprising
that the X-ray luminosity is lower than in Arches, and, in fact,
it may be too large to be explained solely with the cluster
wind scenario.

\section{The LBV HD 5980}

Among all stars in NGC\,346, \hd\ is unique in its spectral
variations in the last two decades, and warrants further examination. 
This eclipsing binary (or triple
system?) was classified as WN+OB before 1980, but its spectral type
changed to WN3+WN4 in 1980--1983, and to WN6, with no trace of the
companion in 1992. In 1994, this star underwent a LBV-type eruption,
and presented at the same time a WN11 type. The eruption has now settled
down and the spectrum is returning to its pre-eruption state (for more
details see Koenigsberger et al. 2000, and references therein). 
We will label the 1994 eruptor as `star A' and its companion as 
`star B'.\\

{\it Chandra} is the first X-ray telescope to detect this peculiar
system, since the low spatial resolution of previous X-ray observatories
did not allow the distinction of \hd\ from the extended emission 
surrounding it.  
However, even with a 100\,ks exposure, the data possess a
rather low signal-to-noise ratio. The {\it Chandra} ACIS-I count rate of 
\hd\ is only $2.98\pm 0.19\times 10^{-3}$~cnts~s$^{-1}$ in the $0.3-
10.0$~keV band or about 300 counts in the total observation. 
This limits our ability to sensitively determine the spectral shape 
of the emission and to look for source variability.  On the other 
hand, the relative faintness of the source means that photon pileup 
will not affect our analysis to any great degree. \\

We have extracted a spectrum of \hd\ using the CIAO tool {\it
psextract}. An annular background region around the source (but 
sufficiently far from it) was chosen, in order to eliminate contamination 
from the surrounding extended emission (see \S~5). This spectrum is 
shown in Fig.~\ref{spec59}a. The best fit model to the spectrum has 
the following parameters: $N(H)=0.22_{0.14}^{0.35}\times 10^{22}$~cm$^{-2}$ 
and $kT=7.04_{4.35}^{13.35}$~keV for an absorbed {\em mekal} model and 
$N(H)=0.28_{0.19}^{0.44}\times 10^{22}$~cm$^{-2}$ 
and $\Gamma$=1.74$_{1.53}^{1.89}$ for an absorbed power-law. 
These absorption columns are consistent with the value expected for 
NGC\,346 (see paper II) but part of these columns are probably due 
to wind absorption and/or absorption from the 1994 ejecta. The derived 
observed X-ray luminosity of \hd\ is $L_X= 1.3\times
10^{34}$~erg~s$^{-1}$ in the $0.3-10.0$~keV band for both models.
Using the normalisation factor of the {\em mekal} model, we found a volume
emission measure of $\sim$10$^{57}$~cm$^{-3}$ for this source.\\

We can compare \hd\ with other WR stars detected in X-rays. 
The unabsorbed X-ray luminosity of \hd\ is $1.7\times 10^{34}$~erg~s$^{-1}$ 
in the $0.3-10.0$~keV energy range and $9\times 10^{33}$~erg~s$^{-1}$ in the
{\it ROSAT} range, i.e. $0.2-2.4$~keV. This places \hd\ amongst the 
X-ray brightest single WN stars \citep{we96} and the brightest WR+OB
binaries \citep{pol} of the Galaxy.\\ 

Two factors could explain this high X-ray luminosity. First, the fast
wind from the post-eruptive phases (from 1300 \kms\ in 1994 to
$\sim$2000 \kms\ in 2000) should now collide with the slow wind (ejected
with a velocity as low as 200 \kms\ during the eruption, see e.g. 
Koenigsberger et al. 2000), which will
increase the post-eruptive X-ray luminosity. X-ray emission from 
colliding ejecta has already been observed for another LBV, i.e. 
$\eta$ Carinae. An elliptically shaped extended emission of 40\arcsec$\times$70\arcsec (i.e. 0.4~pc$\times$0.8~pc) in size 
surrounds the star in the X-ray domain \citep{se01}. It is 
correlated to the high velocity ejecta of the Homunculus 
nebula \citep{we01}, which was created after the last great eruption. 
For \hd, however, the ejecta from the 1994 eruption has not had 
sufficient time to form a detectable LBV nebula around the 
star, and no `Homunculus-like' nebula from a previous eruption is 
visible around \hd\ (see Fig.~\ref{hst6540x}): any X-ray emission 
from the colliding ejecta should thus still be blended with the stellar 
emission from \hd\ and that may contribute to explain the high X-ray 
luminosity. Unfortunately, no X-ray detection of \hd\ before the 1994 
eruption is available. A ROSAT all-sky survey image taken in 1990
(exposure number 933001) provides only an upper limit of 
L$_X\sim$10$^{36}$~erg~s$^{-1}$ 
at the position of \hd. \citet{wa92} gave a 3$\sigma$ upper limit of 2$\times$10$^{34}$~erg~s$^{-1}$ for all SMC WR stars, except AB7.
These limits are compatible with the detected luminosity of \hd\ and 
do not allow us to quantitatively compare the pre- and post-eruption
luminosities to confirm the predicted luminosity enhancement.\\

The fact that \hd\ is a close binary system containing two very massive 
stars, suggests that colliding winds may provide another source for the 
observed X-rays. We can thus estimate the expected X-ray luminosity. 
For the system, we assume that the stellar winds of the stars
have returned to their pre-eruption values and we assume that $\dot
M_A$=1.4$\times$~10$^{-5}$~M$_{\odot}$~yr$^{-1}$, 
$v_\infty(A)$=2500~km~s$^{-1}$ for the O-star, and $\dot M_B$=
2$\times$10$^{-5}$~M$_{\odot}$~yr$^{-1}$, $v_\infty(B)$=1700~km~s$^{-1}$ 
for the WN Wolf-Rayet star \citep{mo98}. Thus the
momenta of each star's wind are very similar, and the total wind kinetic
energy is $5\times 10^{37}$ erg s$^{-1}$. These wind parameters lead to a
momentum ratio of $\eta=(M_B v_\infty(B))/(M_A v_\infty(A))=0.97$ 
\citep{us92} and the wind-wind collision shock will lie halfway between 
the stars. In this configuration we expect about 1/6 of the wind 
kinetic energy to pass perpendicularly through 
the two shocks and be thermalised \citep{st92}, and because the 
systems are relatively close much of this energy will be radiated\footnote{In 
terms of the cooling parameter $\chi$ defined by \citet{st92}, for 
the WN wind $\chi\ll 1$ and for the O-star wind $\chi \sim 1$.}.\\
 
Consequently, we would expect the X-ray luminosity of \hd\ to be $\sim
10^{36}$~erg~s$^{-1}$ or more, rather than the observed value of $\sim
10^{34}$~erg~s$^{-1}$. The discrepancy between the predicted and 
observed X-ray luminosities could be due to a range of factors, e.g. 
the winds may not be at the pre-eruption levels - lower mass-loss rates 
or lower wind velocities will translate to a lower X-ray luminosity. 
Radiative braking, wind clumping, or the eclipse of the colliding 
wind region by the thick wind of star B could also reduce the luminosity.
Alternatively, the wind momenta may not be so nearly equal, as a low value 
of $\eta$ also tends to lower the luminosity. \\

Using the ephemeris of \citet{st97}, we have computed a phase of 
$\phi=0.24-0.30$ for our {\it Chandra} observation: this is close 
to the eclipse of star A by star B ($\phi_{ecl}$=0.36). If the 
eccentricity is $e\sim 0.3$, a simple adiabatic model predicts that 
the X-ray luminosity will vary inversely with separation. We may 
expect a change in the intrinsic X-ray luminosity of a factor of 
$\sim 2$ through the orbit and by $\sim$ 10\% during the {\it Chandra} 
observation. We have detected a variation of apparently larger amplitude 
(factor $\sim$2) in the count rate of \hd\ (see Fig. \ref{spec59}b). 
This still suggests we are seeing orbital variability associated with 
colliding winds, since the additional effects of changing 
absorption\footnote{The colliding wind region is viewed alternatively
through the atmosphere of stars A and B. The absorption thus changes 
with the orbital phase.}
on the observed luminosity can either reduce or enhance this variability. 
This observed variability is much more likely to be due to 
colliding winds than wind-blown bubble type emission, where no short 
timescale variability would be expected. The high fitted temperature 
of \hd\ $kT\sim 7$~keV also suggests that we are likely seeing colliding wind 
emission, as this temperature corresponds to a shock velocity of 
$\sim 2500$ km s$^{-1}$, comparable to the wind speed of star A 
($v_\infty(A)$ above).\\

The detailed characteristics of the X-ray properties of \hd\ 
need to be studied further, in conjunction with detailed colliding wind 
models (c.f. the case of $\eta$ Carinae; Pittard \& Corcoran 2002). An 
{\it XMM-Newton} observation is scheduled at phases $\phi=0.09-0.10$, 
just after periastron: the comparison between these two datasets may enable 
us to detect further the signature of the colliding wind region, and 
constrain it more precisely.\\

Finally, we note that the unabsorbed X-ray luminosity of the central 
source in $\eta$ Carinae, if we assume a distance of 2.3~kpc, varies 
between (0.6 and 2.5)$\times$10$^{35}$~erg~s$^{-1}$ \citep{is99}.
Such luminosities are much higher than for \hd.

\section{The Extended Emission around \hd}

Around the position of \hd\ lies a region of bright extended X-ray 
emission. It was first detected by the {\it Einstein Observatory}
(source IKT 18 in Inoue, Koyama, \& Tanaka 1983), and subsequently 
observed by {\it ROSAT} (source [HFP2000] 148\footnote{This {\it ROSAT} 
source may encompass some of the X-ray emission from the cluster.} in 
Haberl et al. 2000). The {\it Chandra} image reveals the extended 
emission in much more detail than previous X-ray observatories. The
overall shape of the emission region is more or less rectangular, with an
extension to the northeast. It contains a few bright or dark arcs, but
apart from these, the brightness is rather uniform, with no obvious
limb-brightening. It has a size of $130\arcsec\times 100\arcsec$,
i.e. 37pc$\times$29pc at the SMC's distance. \hd\ lies towards the top
center of the emission region. An X-ray bright filament extends from 9\arcsec\
west to 23\arcsec\ south of \hd.  An X-ray dark feature
appears some 30\arcsec\ at the southeast of \hd.  The total count rate
of this source in the $0.3-10.0$~keV energy range is $7.25\pm 0.17\times
10^{-2}$~cnts~s$^{-1}$. It is the softest source present in the field.\\

The spectrum of this extended X-ray emission was extracted in a rectangular aperture of $\sim$160\arcsec$\times$110\arcsec\ with a carefully chosen 
background region, located as close as possible to the source and in a
region on the ACIS-I1 CCD where there are no point sources.
A circular region of diameter $\sim$5\arcsec\ containing \hd\ was 
removed from the extraction.
As the X-ray emission is extended the reponse matrices were calculated 
using the {\it calcrmf} and {\it calcarf} tools.  The best fit 
($\chi^2_{\nu}$=1.03, $N_{dof}$=244) to this spectrum was an absorbed 
{\em mekal} model with the following properties (see Fig.~\ref{specext}):
$N(H)=0.12_{0.10}^{0.15}\times 10^{22}$~cm$^{-2}$,
$kT=0.66_{0.64}^{0.68}$~keV and $Z=0.17_{0.15}^{0.21} Z_{\odot}$. The
observed flux in the $0.3-10.0$~keV band was $3.35\times 10^{-13}$~erg~cm
$^{-2}$ s$^{-1}$, i.e. an observed luminosity of $1.40\times 
10^{35}$~erg~s$^{-1}$. The low value of the absorption column, compared
to \hd\ and the cluster, suggests that this X-ray source 
lies between NGC\,346 and the observer, but still in the SMC (see
paper II).\\

Using the normalisation factor of the {\em mekal} model, we found a volume
emission measure $\int n_en_H dV$ of $\sim$3$\times$10$^{58}$~cm$^{-3}$.
Considering a constant density for the hot gas, a sperical geometry
of diameter $\sim$33~pc for the extended emission and assuming a 
pure H composition, the density of the hot gas is roughly 0.2~cm$^{-3}$ 
and the total mass of the hot gas is $\sim$100~M$_{\odot}$. \\

Using the available data, a deeper analysis was conducted, searching for
the existence of a temperature gradient throughout the source area, with
color and temperature maps. For color maps, we generated a set of 
narrow-band images, e.g. $0.3-0.9$~keV, $0.9-1.2$~keV and $1.2-2.0$~keV, 
that enable us to 
distinguish the harder components from the softer parts. In addition,
temperature maps were constructed from spectral fits in small regions
of the SNR.  The only obvious trend in both the color and temperature
maps is the softness of the northeastern extension of the SNR. However,
it is not possible to draw any firm conclusions regarding the presence
of any temperature gradient.\\

We have also constructed images of the extended emission in narrow 
energy bands, each correspond to a specific ion. The energy bands 
used for each ion are shown in Fig.~\ref{ion}. The data in 
each band were binned to obtain $4.9\arcsec
\times 4.9\arcsec$\ pixels (see Fig.~\ref{ion}). In contrast to
N132D \citep{beh}, no significative differences between the morphology
of highly ionized species (Mg$^{10+}$, Ne$^{9+}$, Si$^{12+}$, Fe$^{19+}$)
and lower ionization species were detected.\\

\subsection{The Nature of the Extended Emission}

Since its discovery the extended X-ray emission has been attributed to
a supernova remnant (SNR). Further evidence was subsequently obtained
to support this hypothesis; for example, \citet{ye91} found a
non-thermal radio source at this position that they called
SNR\,$0057-7226$. Using the lowest contours of \citet{ye91}, the radio
source is 56~pc$\times$52~pc, slightly larger than the X-ray source.\\

Moreover, evidence of high velocity motions in this region were
observed in the visible and UV ranges. Using echelle spectra centered on
\ha, \citet{ch88} found clumps moving at v$_{LSR}\sim$ 300\kms, 
redshifted from the main quiescent component by +170\kms: only the 
receding side of
the expanding object is seen, suggesting a low density where the
approaching side arises. UV analyses with {\it IUE} \citep{de80}, {\it
HST} STIS \citep{ko01} and {\it FUSE} \citep{ho01} have also confirmed
the presence of an expanding structure.  While \citet{de80} and
\citet{ho01} found only absorptions at v$_{LSR}$=+300\kms, \citet{ko01}
detected both expanding sides of the object, with components at
$v_{LSR}= +21, +52, +300, +331 \kms$. However, the absorptions at +21
and +52 \kms\ were small, and need a future confirmation. The presence
of components from both approaching and receding sides of the expanding 
structure has now been confirmed by \citet{da02} from deep echelle
spectroscopic integrations at optical wavelengths of the N66 region. 
Absorptions at any of these
velocities are not seen in the spectra of Sk~80, a close neighbor of
\hd\ situated on the edge of the X-ray source. A fast expanding
structure close to the position of \hd\ is thus present.\\

Unfortunately, even the high-resolution {\it HST} WFPC2 images of the
close neighborhood of \hd\ do not show any clear nebula associated
with the X-ray extended emission (see Fig.~\ref{hst6540x}). A lot of
filamentary structures - generally indicative of a SNR \citep{ch00} -
are seen throughout the field, but are not limited to the exact location
of the X-ray source. There is thus no obvious \ha\ feature directly
correlated with the extended X-ray emission. \\

Considering all the evidence (non-thermal emission, high velocity
expanding shell etc), the X-ray / radio source should thus be regarded
as a SNR located in front of \hd\ but still belonging to the SMC. The SNR
hypothesis is further supported by the comparison of the size of this
feature to its X-ray and radio fluxes \citep{ma83}. \\

As \hd\ is projected more or less at the center of the diffuse
X-ray emission and it underwent a LBV-type eruption in 1994 (which is
unlikely to be the only one undergone by the system), it is
tempting to associate the diffuse X-ray emission with \hd.
To investigate whether this association is likely, we first compare \hd\
with $\eta$ Carinae, the most well-known LBV that have gone through
multiple eruptions.  $\eta$ Car is also surrounded by bright, extended
X-ray emission that appears to be assciated with the Carina Nebula:
it is comparable in X-ray temperature ($kT$), size and morphology 
to the X-ray emission around \hd\ (Seward \& Mitchell 1981; 
Fig.\ref{carina}). But it is generally assumed that this Nebula
has been formed through the collective action of all stars of the Carina
cluster, not only by the single action of $\eta$ Carinae.\\

We have also searched for clues of an interaction between \hd\ and 
its surroundings (e.g. a LBV nebula). \citet{wa78} noted evidence of 
such interaction: an arc of radius $\sim 30\arcsec$
centered on the star is actually visible in \ha\ and \oiii\ (but not in
\sii). However, this arc is rather diffuse (see Fig.~\ref{hst6540x}),
indicating a lack of shock compression. Moreover, its velocity 
coincides with that of the quiescent region, and no line-splitting 
region is seen beginning at the arc and extending towards the star 
\citep{da02}: this arc is certainly not the rim of 
an expanding shell. This optical arc has different size, shape, 
and location from an X-ray bright arc (Fig.~\ref{hst6540x}), therefore 
we conclude that they are probably two distinct features unrelated 
to each other. Echelle spectra of this area also do not show any 
enhancement of the \nii/\ha\ ratio which usually indicates the presence of 
circumstellar material surrounding the star. The comparison between 
X-rays and visible data thus do not show any clear
indication of an interaction between \hd\ and its environment.
We also note that no circumstellar nebula produced by \hd\ could 
explain the redshifted UV absorptions in the spectra of the star, or 
the non-thermal radio emission from the extended source. \\

The last possibility is that the source is indeed a SNR, but that \hd\
is still directly responsible for it. Several authors have proposed the
existence of a third component in \hd, `star C', to interpret the visual
light curve \citep{br91} and the presence of stable photospheric 
absorptions in UV 
\citep{ko00}. To explain the existence of a SNR, an additional, fourth 
component of the system, `star D', is needed. It would now be a 
compact object, a remnant of the star that exploded in SN long ago. 
If star C is much farther away than the system A+B+D, then the high velocity 
redshifted UV absorptions could be produced by star C shining through 
the receding back wall of the SNR. The presence of a compact object
might explain the $\sim$6~h variations seen in the lightcurve of \hd\
by \citet{st97}. Accretion onto this compact object might also constitute 
another source of X-rays, thus contributing to understand the high X-ray 
emission from the system.\\

\section{Summary}

In this series of articles, we report the analysis of the {\it Chandra} data
of N66, the largest star formation region of the SMC. 
In this first paper, we have focused on the most important 
objects of the field: the NGC\,346 cluster and \hd. \\

The cluster itself is relatively faint, with a total luminosity of 
$L_X^{unabs}\sim 1.5\times 10^{34}$~erg~s$^{-1}$ in the 0.3-10.0~keV 
energy range. Most of this emission seems correlated
with the location of the brightest stars of the core of the cluster, but
the level of X-ray emission probably cannot be explained solely by the
emission from individual stars. \\

In this field lies another object of interest: \hd, a remarkable star
that underwent a LBV-type eruption in 1994. {\it Chandra} is in fact the
first X-ray telescope to detect \hd\ individually. In X-rays, the star
appears very bright, comparable only to the brightest WR stars in the
Galaxy, but still fainter than $\eta$ Carinae.  The comparison of our 
results with future X-ray observations
will enable us to better understand \hd: for example, phase-locked
variations will be analysed in the perspective of the colliding wind
behaviour of this binary, while other variations may be related to the
recent LBV eruption.  Follow-up observations are thus needed to complete
the study of this system. \\

A bright, extended X-ray emission is seen to
surround \hd. It is most probably due to a SNR whose progenitor 
is unknown. The spatial coincidence of this extended X-ray emission
with the peculiar massive star suggest an association between these
two objects. We also note the close resemblance of this X-ray emission
to the Carina Nebula in which the LBV $\eta$ Carinae lies. \\

\acknowledgments

We are grateful to Dr Martin A. Guerrero Roncel for useful 
suggestions on data analysis techniques.
Support for this work was provided by the National
Aeronautics and Space Administration through Chandra Award Number
GO1-2013Z issued by the Chandra X-Ray Observatory Center, which is
operated by the Smithsonian Astrophysical Observatory for and on
behalf of NASA under contract NAS8-39073.
Y.N. acknowledges support from the PRODEX XMM-OM and Integral Projects, 
contracts P4/05 and P5/36 `P\^ole d'attraction Interuniversitaire 
(SSTC-Belgium) and from PPARC for an extended visit to the University 
of Birmingham. IRS and JMH also acknowledge support from PPARC.  
AFJM thanks NSERC (Canada) and FCAR (Quebec) for financial aid.
This research has made use of the 
SIMBAD databse, operated at CDS, Strasbourg, France and NASA's Astrophysics 
Data System Abstract Service.

\clearpage

\begin{figure}
%\plotone{f1.eps}
\caption{{\it Chandra} color image of the region close to \hd. 
Three energy bands were used to create this color image : red corresponds
to 0.3-1.0~keV, green to 1.0-2.0~keV and blue to 2.0-10.0~keV.
Before combination, the images were smoothed by convolution with 
a gaussian of $\sigma$=1\arcsec.
\label{chandracol}}
\end{figure}

\begin{figure}
%\plotone{f2.ps}
\caption{{\it HST} WFPC2 \ha\ data of the region surrounding \hd\
(which is the star located at the center of the PC chip). The X-ray
data overlaid as contours on the {\it HST} image were first binned 
by a factor of 2 and then smoothed by convolution with a gaussian of 
$\sigma$=2\arcsec. 
\label{hst6540x}}
\end{figure}

\begin{figure}
%\plotone{f3.ps}
\caption{The {\it Chandra} X-ray data of the NGC\,346/\hd\
region superimposed on the {\it DSS} image, showing the extended 
nature of the X-ray source associated with the cluster. The location 
of the brightest stars in the cluster are indicated.  The X-ray data 
were first binned by a factor of 2 and then smoothed by 
convolution with a gaussian of $\sigma=2\arcsec$)
\label{dssmpg}}
\end{figure}

\begin{figure}
%\plotone{f4.ps}
\caption{The X-ray spectrum of the NGC\,346 cluster, shown
along with the best-fit absorbed {\em mekal} model, with $N(H)=0.42\times
10^{22}$~cm$^{-2}$ and $kT=1.01$~keV (see \S~3).\label{specclus}} 
\end{figure}

\begin{figure}
%\plottwo{f5a.ps}{f5b.ps}
\caption{a. The {\it Chandra} X-ray spectrum of \hd, shown along with the
best-fit absorbed {\em mekal} model, with $N(H)=0.22\times 10^{22}$~cm$^{-2}$ 
and $kT=7.04$~keV. b. The {\it Chandra} X-ray lightcurve of \hd, in the 
0.3-10~keV range and with 10 bins of 10~ks each.
\label{spec59}} 
\end{figure}

\begin{figure}
%\plotone{f6.ps}
\caption{The {\it Chandra} X-ray  spectrum of the extended emission
around \hd, shown along with the best-fit
absorbed {\em mekal} model, with $N(H)=0.12\times 10^{22}$~cm$^{-2}$, 
$kT=0.66$~keV and $Z=0.17Z_\odot$ (see \S~5).
\label{specext}}
\end{figure}

\begin{figure}
%\plotone{f7.ps}
\caption{Narrow-band X-ray images of the extended emission surrounding
\hd. Each image is labeled with the principal line-emitting ion and the
energy range used to generate the image. 
\label{ion}}
\end{figure}

\begin{figure}
%\plotone{f8.ps}
\caption{ X-ray images of the Carina Nebula (24~ks ; ROSAT PSPC 
observation \#900176) compared to the extended emission around 
\hd. If we assume a distance of 2.3~kpc for the Carina 
Nebula and 59~kpc for the SMC, both images are $\sim$55~pc$\times$55~pc. 
ROSAT PSPC support rings and {\it Chandra} ACIS-I CCD gaps can be spotted 
on these images. Both images were smoothed by convolution with a gaussian
($\sigma$=15\arcsec\ for the Carina Nebula and $\sigma$=2\arcsec\ after 
binning by a factor of 2 for \hd).
\label{carina}}
\end{figure}

%\clearpage 

\end{document}